\documentclass[11pt,a4paper,twoside]{article}
\usepackage{pstricks,pst-node} 
\usepackage{amsmath,amssymb}
\usepackage{times,cite}

\newtheorem{theorem}{Theorem}[section]
\newtheorem{corollary}{Corollary}[section]

\title{\Huge Bidifferential graded algebras \\ and integrable systems}

\newcommand{\be}{\begin{eqnarray}}
\newcommand{\ee}{\end{eqnarray}}
\newcommand{\bez}{\begin{eqnarray*}}
\newcommand{\eez}{\end{eqnarray*}}
\newcommand{\pa}{\partial}
\newcommand{\la}{\lambda}
\newcommand{\bt}{\mathbf{t}}
\renewcommand{\d}{\mathrm{d}}
\newcommand{\bd}{\bar{\mathrm{d}}}
\newcommand{\D}{\mathrm{D}}
\newcommand{\bD}{\bar{\mathrm{D}}}
\newcommand{\cA}{\mathcal{A}}
\newcommand{\E}{\mathbb{E}}

\author{ {\scshape Aristophanes Dimakis} \\
 Department of Financial and Management Engineering, \\
 University of the Aegean, 31 Fostini Str., GR-82100 Chios, Greece \\
 E-mail: \emph{dimakis@aegean.gr}
          \and
 {\scshape Folkert M\"uller-Hoissen} \\
 Max-Planck-Institute for Dynamics and Self-Organization \\
 Bunsenstrasse 10, D-37073 G\"ottingen, Germany \\
 E-mail: \emph{folkert.mueller-hoissen@ds.mpg.de} 
       }

\date{}

\begin{document}

\maketitle

\begin{abstract}
In the framework of bidifferential graded algebras, we present 
universal solution generating techniques for a wide class of 
integrable systems. 
\end{abstract}

\section{Introduction}
\label{section:intro}
\setcounter{equation}{0}
Let $\cA$ be a unital associative algebra (over $\mathbb{R}$ or $\mathbb{C}$)
with identity element $I$, and 
$\Omega(\cA) = \bigoplus_{r \geq 0} \Omega^r(\cA)$ with $\Omega^0(\cA) = \cA$ 
and $\cA$-bimodules $\Omega^r(\cA)$, $r=1,2,\ldots$. 
We call $(\Omega(\cA), \d, \bd)$ a \emph{bidifferential graded algebra} (BDGA),
or \emph{bidifferential calculus}, 
if $(\Omega(\cA), \d)$ and $(\Omega(\cA), \bd)$ are both differential graded 
algebras, which means that $\Omega(\cA)$ is a graded algebra and the linear maps 
$\d,\bd : \Omega^r(\cA) \rightarrow \Omega^{r+1}(\cA)$ satisfy the graded 
Leibniz rule (antiderivation property) and 
\be 
    \d^2 = \bd^2 = 0  \, , \qquad  
    \d \, \bd + \bd \, \d = 0 \; .
           \label{bidiff-cond}
\ee
These conditions can be combined into $\d_z^2 = 0$, where 
$\d_z := \bd - z \, \d$ with an indeterminate $z$.\footnote{A generalization 
to an $N$-differential graded algebra is then obtained if $\d_z^2=0$ with 
$\d_z = \sum_{n=0}^N z^n \, \d_n$. But this will not be considered in this work.} 
In section~\ref{section:dress} we connect this structure with 
`integrable' partial differential (or difference) equations 
\cite{DMH00a,DMH00b,DMH00c,DMH00d,DMH00e,DMH01bt,DMH01FK,DMH02LV,Ma+Shi00,Sita00,CST00a,CST00b,Cram+Sarl01bi,Gueu01,Lega02,Cama+Cari03,Chav03,Chav05,Gris+Pena03,Zuev05,Lorenzoni06}. 
Although this framework may not be able to cover all possible (in some sense) 
integrable equations, it has the advantage of admitting \emph{universal} 
techniques for constructing exact solutions. Whereas previous work 
concentrated on conservation laws and B\"acklund transformations, the present 
work addresses Darboux transformations and presents a very effective 
`linearization approach', generalizing results in \cite{DMH07disp} 
(see also \cite{March88,Carl+Schi99} for related ideas).
After collecting some basics in section~\ref{section:dress}, 
section~\ref{section:sgt} addresses universal solution generating techniques. 
Section~\ref{section:examples} then presents some examples. 
Section~\ref{section:conclusions} contains final remarks.

\section{Dressing bidifferential graded algebras}
\label{section:dress}
\setcounter{equation}{0}
In the following, $(\Omega(\cA), \d, \bd)$ denotes a BDGA. Introducing
\be
     \bD := \bd - A
\ee
with a 1-form $A$, $\d$ and $\bD$ satisfy again the BDGA relations iff
\be
    \bd A - A \, A \equiv -\bD^2 = 0 \qquad \mbox{and} \qquad
    \d A \equiv - (\d \, \bD + \bD \, \d) = 0 \; . \label{gauged-bdga-cond}
\ee
We are interested in cases where these equations are equivalent to a 
partial differential or difference equation (or a family of such equations), 
which requires that $A$ depends on a set of independent variables and 
the differential maps $\d, \bd$ involve differential or difference operators. 
As depicted in the following diagram, we can solve either the first or 
the second equation. 
\bez
 \setlength{\arraycolsep}{1.8cm}
 \begin{array}{cc}
   \Rnode{a}{A = (\bd g) \, g^{-1}} & 
       \Rnode{b}{ \d \, \Big( (\bd g) \, g^{-1}\Big) =0 } \\[.8cm]
   \Rnode{c}{ \bd A - A \, A = 0 } 
      & \Rnode{d}{\d A = 0 } \\[.7cm]
   \Rnode{e}{ \bd \, \d \, \phi = \d\phi \; \d\phi } 
        & \Rnode{f}{ A = \d \phi } 
 \end{array}
 \psset{nodesep=5pt,arrows=->,linestyle=solid}
 \ncLine{a}{b}
 \ncLine{f}{e}
 \psset{nodesep=5pt,arrows=<->,linestyle=dashed}
 \ncLine{c}{e}
 \ncLine{d}{b}
 \psset{nodesep=5pt,arrows=->,linestyle=solid,linecolor=red}
 \ncLine{c}{a} 
 \ncLine{d}{f} 
 \psset{nodesep=5pt,arrows=<->,linestyle=dashed,linecolor=green}
 \ncline{b}{e}\Bput{\mbox{`pseudoduality'}\hspace{-2cm}}
\eez
This results in two different equations that are related by a 
`Miura transformation' 
\be
     (\bd g) \, g^{-1} = \d \, \phi \, ,  \label{Miura}
\ee
and this relationship is sometimes referred to as `pseudoduality'. 

The conditions (\ref{gauged-bdga-cond}) can be combined into 
\be
   \D_z^2 = 0 \qquad \mbox{where} \qquad  
   \D_z = \bD - z \, \d = \d_z - A \; . 
\ee
Such a zero curvature condition is at the roots of the theory of 
integrable systems. It is the integrability condition of the linear equation
\be
   \D_z W(z) = 0 \; .   \label{D_zW(z)=0}
\ee
To get some more information about this equation, let us derive it from 
\be
    \bd \, \d \, \phi = \d \phi \; \d \phi  \, ,    \label{univ_eq}
\ee
the equation in the lower left corner of the diagram. 
Using (\ref{bidiff-cond}), we write it as
\be
    \d [ \bd \phi - (\d \phi) \, \phi ] = 0 \; .
\ee
We shall assume that the first $\d$-cohomology class vanishes, so that 
$\d$-closed 1-forms are $\d$-exact.
Then there is an element $\psi \in \cA$ such that
\be
    \bd \phi - (\d \phi) \, \phi = \d \psi \; .  \label{bdphi-psi}
\ee
Applying $\bd$, using (\ref{bidiff-cond}), (\ref{univ_eq}) and 
(\ref{bdphi-psi}),  we obtain
\be
    \d [ \bd \psi - (\d \phi) \, \psi ] = 0 \, , 
\ee
which in turn can be integrated by introduction of a new potential $\chi \in \cA$, 
\be
    \bd\psi - (\d\phi) \, \psi = \d \chi \; .    \label{bdpsi-chi}
\ee
This procedure can be iterated and yields the linear equation (\ref{D_zW(z)=0}) 
with 
\be
    W(z) = I + \sum_{n \geq 1} W_n \, z^{-n} \, , \quad W_1 = -\phi \; . 
\ee
\textit{Remark.} 
A gauge transformation of a BDGA $(\Omega(\cA), \d, \bd)$ 
is given by $\d \mapsto G \, \d \, G^{-1} = \d'$, 
$\bd \mapsto G \, \bd \, G^{-1} = \bd'$ with 
an invertible map $G : \cA \rightarrow \cA$. 
In case of $(\Omega(\cA), \d, \bD)$, choosing $G$ as multiplication 
by $g^{-1}$, we obtain the equivalent BDGA $(\Omega(\cA), \d', \bD')$
with $\d' = \d + g^{-1} \, \d g$ and $\bD' = \bd$, by use of 
$A = (\bd g) \, g^{-1}$. 
\vskip.1cm

\noindent
\textit{Remark.} 
We can consider a simultaneous dressing of $\d$ and $\bd$  
by introducing $\D = \d - B$ in addition to $\bD = \bd - A$. 
Then $(\Omega(\cA), \D, \bD)$ is a BDGA iff
\be
    \d B = B \, B \, , \qquad \bd A = A \, A \, , \qquad
    \d A + \bd B = A \, B + B \, A \; .
\ee
Solving the first two conditions by setting $A = (\bd g) \, g^{-1}$, 
$B = (\d h) \, h^{-1}$, the third (multiplied from the left by $h^{-1}$,  
from the right by $h$) becomes 
$\d [ (\bd J) \, J^{-1}] = 0$ with $J = h^{-1} g$.
An equivalent form is $\bd [ (\d J^{-1}) \, J] = 0$.
This generalizes Yang's gauge in the case of the (anti-) 
self-dual Yang-Mills equation (see also \cite{Taka01}).

\section{Solution generating techniques}
\label{section:sgt}
\setcounter{equation}{0}

\subsection{B\"acklund transformation (BT)}
An elementary BT is given by 
\be
    \D'_z = G(z) \, \D_z \, G(z)^{-1} \, , \qquad 
    \D'_z := \d_z - A' \, , 
\ee
where $G(z) = I + F \, z^{-1}$ \cite{DMH01bt}. 
This is equivalent to
\be
    \d F = A - A' \, , \qquad  
    \bd F = A' \, F - F \, A \; .
   \label{BT-F_eqs}
\ee
Using $A = \d \phi$, the first equation can be integrated, 
\be
    F = \phi-\phi' - C \qquad \mbox{where} \qquad \d C =0 \, , 
\ee
and from the second equation we obtain the elementary BT
\be
  \bd (\phi' - \phi + C) = (\d \phi') \, (\phi' - \phi +C) 
                           - (\phi'-\phi +C) \, \d \phi \, .  \label{BT}
\ee
Alternatively, using $A = (\bd g) \, g^{-1}$, the second of 
equations (\ref{BT-F_eqs}) is solved by 
\be
    F = g' \, \mathcal{K} \, g^{-1} \qquad \mbox{where} \qquad
    \bd \mathcal{K} = 0 \, ,
\ee
and the first of equations (\ref{BT-F_eqs}) becomes
\be
   \d(g' \mathcal{K} g^{-1}) = (\bd g') \, g'{}^{-1} - (\bd g) \, g^{-1}
    \, ,  \label{BT-g}
\ee
which is the elementary BT for the pseudodual equation. Using the 
Miura transformation (\ref{Miura}), this can be integrated and yields
\be
    \phi'-\phi + C = g' \, \mathcal{K} \, g^{-1}  \; .    \label{phi-g-BT}
\ee
This equation connects the two elementary BTs.

\subsection{Darboux transformation (DT)}
\label{subsec:Darboux}
The linear system\footnote{Instead of (\ref{bdpsi_Delta}), we may consider 
$\bd \psi = (\d \phi) \; \psi + \d (\psi \, \Delta)$, which results 
from (\ref{bdpsi-chi}) by setting $\chi = \psi \, \Delta$. In this 
case we have to impose $\bd \Delta = \Delta \, \d \Delta$ in order to 
obtain (\ref{univ_eq}) as integrability condition. Some of the 
following formulae, also in section~\ref{subsec:binDarboux}, then 
have to be modified accordingly. One can prove that the two possibilities 
are in fact equivalent.}
\be
    \bd \psi = (\d \phi) \; \psi + (\d \psi) \, \Delta \, , 
               \label{bdpsi_Delta}
\ee
has the following integrability condition,
\be
    \Big( \bd \d \phi - (\d \phi)^2 \Big) \, \psi
  - \d \Big( \psi \, ( \bd \Delta - (\d \Delta) \, \Delta) \Big) = 0 \, , 
\ee 
which reduces to (\ref{univ_eq}) if 
\be
    \bd \Delta = (\d \Delta) \, \Delta \; . \label{bdDelta}
\ee
Let $\theta$ be an invertible solution of (\ref{bdpsi_Delta}) with 
a solution $\Delta'$ of (\ref{bdDelta}), hence
\be
    \bd \theta = (\d \phi) \; \theta + (\d \theta) \, \Delta' \; .
         \label{Darb_bdtheta}
\ee
As a consequence, 
\be
    \bd(\theta \, \Delta' \, \theta^{-1}) 
 = (\d \phi') \; \theta \, \Delta' \, \theta^{-1} 
    - \theta \, \Delta' \, \theta^{-1} \, \d \phi \, , 
\ee
where 
\be
    \phi' := \phi + \theta \, \Delta' \, \theta^{-1} - C'  
    \qquad \mbox{with} \qquad \d C' =0 \; .
             \label{Darb_phi'}
\ee
This is in accordance with (\ref{BT}), i.e. $\phi'$ is related to $\phi$ 
by an elementary BT. 
Hence, any solution $\phi$ of (\ref{univ_eq}) and any invertible solution 
$\theta$ of the linear equation (\ref{Darb_bdtheta}) determine a new 
solution $\phi'$ of (\ref{univ_eq}) via (\ref{Darb_phi'}). This is an 
abstraction of what is known as a Darboux transformation (see e.g. \cite{Matv+Sall91}). 
Introducing
\be
    \psi' = (\psi \, \Delta - \theta \, \Delta' \, \theta^{-1} \psi) 
    \, \mathcal{M} \, ,  \label{Darb_psi'}
\ee
where $\mathcal{M}$ satisfies
\be
    \bd \mathcal{M} = (\d \mathcal{M}) \, \Delta  \, , \qquad
    [\Delta,\mathcal{M}]=0 \, ,   \label{Darboux_M-conds}
\ee
it follows that $\psi'$ satisfies (\ref{bdpsi_Delta}) with $\phi$ replaced 
by $\phi'$, i.e.
\be
    \bd \psi' = (\d \phi') \; \psi' + (\d \psi') \, \Delta \; .
\ee
Now we can iterate this procedure. 
Let $\theta_k$, $k=1,\ldots,n$, be invertible solutions of 
$\bd \theta_k = (\d \phi) \, \theta_k + (\d \theta_k) \, \Delta_k$, 
and $\mathcal{M}_k$ satisfy (\ref{Darboux_M-conds}) with $\Delta_k$. 
Set $\psi_{[1]} = \psi$, $\theta_{[1]}=\theta_1$, 
\be
    \psi_{[k+1]} = (\psi_{[k]}\Delta 
  - \theta_{[k]} \, \Delta_k \, \theta_{[k]}^{-1} \, \psi_{[k]}) \, \mathcal{M}  
  \quad \mbox{with} \quad
  \theta_{[k]} = \psi_{[k]} \Big|_{\psi \to \theta_k,\; \Delta \to \Delta_k,
                 \, \mathcal{M} \to \mathcal{M}_k}  \;
\ee
Then $\psi_{[n+1]}$ satisfies  
$\bd \psi_{[n+1]} = (\d \phi_{[n+1]}) \, \psi_{[n+1]} + (\d \psi_{[n+1]}) \, \Delta$
with the following solution of (\ref{univ_eq}), 
\be
 \phi_{[n+1]} 
 = \phi + \sum_{k=1}^n (\theta_{[k]} \, \Delta_k \, \theta_{[k]}^{-1} - C_k) \; .
\ee
If $\psi_{[n+1]}$ is invertible, then it solves (\ref{mpd}) below (see 
the next subsection).

\subsection{Modified Miura transformation}
If $\psi$ in (\ref{bdpsi_Delta}) is invertible, we have
\be
    [ \, \bd g - (\d g) \, \Delta \, ] \, g^{-1} = \d \, \phi \, , 
    \label{mMiura}
\ee
writing $g$ instead of $\psi$. The integrability condition is 
\be
    \d ( [ \, \bd g - (\d g) \, \Delta \, ] \, g^{-1} ) = 0 \, ,  
            \label{mpd}
\ee
a \emph{modified pseudodual} of (\ref{univ_eq}), related by the 
\emph{modified Miura transformation} (\ref{mMiura}).\footnote{If 
$\Delta = \la \, I$ with $\la \in \mathbb{C}$, the modification can be 
absorbed by a redefinition $\bd' := \bd - \la \, \d$ of $\bd$. }
(\ref{mpd}) corresponds to 
\be
    A = [ \, \bd g - (\d g) \, \Delta \, ] \, g^{-1} \, , 
\ee
which reduces the two equations (\ref{gauged-bdga-cond}) 
to a single one since $\bd A - A \, A = (\d A) \, g \, \Delta \, g^{-1}$. 
We note that (\ref{mMiura}) is equivalent to
\be
  \bd g^{-1} + g^{-1} \, \d \phi' = \d (\Delta \, g^{-1}) 
   \qquad \mbox{where} \qquad 
  \phi' = \phi + g \, \Delta \, g^{-1}
     \; . \label{mMiura2} 
\ee

\subsection{Binary Darboux transformation (bDT)}
\label{subsec:binDarboux}
(\ref{univ_eq}) is also integrability condition of
\be
  \bd\tilde{\psi} = -\tilde{\psi} \, \d\phi + \tilde{\Delta} \, \d\tilde{\psi} 
    \qquad \mbox{where} \qquad 
  \bd\tilde{\Delta} = \tilde{\Delta} \, \d\tilde{\Delta} \; .
      \label{bdtpsi_Delta}
\ee
Combining this with (\ref{bdpsi_Delta}), we get
\be
    \bd(\tilde{\psi} \, \psi)  
  = \tilde{\Delta} \, (\d \tilde{\psi}) \, \psi
    + \tilde{\psi} \, (\d \psi) \, \Delta \; . 
\ee
Introducing $\Omega(\tilde{\psi},\psi)$ via
\be
    \tilde{\Delta} \, \Omega(\tilde{\psi},\psi) 
  - \Omega(\tilde{\psi},\psi) \, \Delta 
  = \tilde{\psi} \, \psi \, ,   \label{Omega}
\ee
the previous equation is satisfied if
\be
    \bd \Omega(\tilde{\psi},\psi) = (\d\Omega(\tilde{\psi},\psi)) \, \Delta
      - (\d \tilde{\Delta}) \, \Omega(\tilde{\psi},\psi) 
      + (\d\tilde{\psi}) \, \psi \; .   \label{bdOmega}
\ee
Now let $\Theta = (\theta_1,\ldots,\theta_N)$ and 
$\tilde{\Theta} = (\tilde{\theta}_1,\ldots,\tilde{\theta}_N)^T$
be solutions of 
\be
    \bd \Theta = (\d\phi) \, \Theta + (\d\Theta) \, \mathbf{\Delta} \, ,
    \qquad
    \bd \tilde{\Theta} = -\tilde{\Theta} \, \d\phi 
           + \tilde{\mathbf{\Delta}} \, \d \tilde{\Theta} \, ,
    \label{bdTheta,bdtTheta}
\ee
with matrices $\mathbf{\Delta},\tilde{\mathbf{\Delta}}$ where 
$\bd\mathbf{\Delta} = (\d\mathbf{\Delta}) \, \mathbf{\Delta}$,  
$\bd\tilde{\mathbf{\Delta}} = \tilde{\mathbf{\Delta}} \, \d\tilde{\mathbf{\Delta}}$, and $\mathbf{\Omega}$ a matrix such that 
\be
 \tilde{\mathbf{\Delta}} \, \mathbf{\Omega} - \mathbf{\Omega} \, \mathbf{\Delta} 
  = \tilde{\Theta} \, \Theta  
    \qquad \mbox{and} \qquad
 \bd\mathbf{\Omega} 
 = (\d\mathbf{\Omega}) \, \mathbf{\Delta} 
   - (\d\tilde{\mathbf{\Delta}}) \, \mathbf{\Omega} 
   + (\d\tilde{\Theta}) \, \Theta     \label{Theta_conds}
\ee
(or equivalently $\bd\mathbf{\Omega} = \tilde{\mathbf{\Delta}} \, 
\d \mathbf{\Omega} - \mathbf{\Omega} \, \d \mathbf{\Delta} - \tilde{\Theta} 
\, \d\Theta$) holds.\footnote{We note that the first of equations 
(\ref{Theta_conds}) is a rank one condition.} 
It follows that
\be
    \Theta' = \Theta \, \mathbf{\Omega}^{-1} \tilde{\mathbf{N}}^{-1} \, , 
      \qquad
    \tilde{\Theta}' = \mathbf{N}^{-1} \mathbf{\Omega}^{-1} \, \tilde{\Theta} 
    \, ,
\ee
satisfy (\ref{bdTheta,bdtTheta}) with $\mathbf{\Delta}$ and 
$\tilde{\mathbf{\Delta}}$ exchanged and $\phi$ replaced by 
\be
   \phi' 
 = \phi - \Theta \, \mathbf{\Omega}^{-1} \tilde{\Theta}
 = \phi-\Theta' \tilde{\mathbf{N}} \, \tilde{\Theta} 
 = \phi - \Theta \, \mathbf{N} \, \tilde{\Theta}' \, , 
   \label{bDT-phi'}
\ee 
if the matrices $\mathbf{N},\tilde{\mathbf{N}}$ are invertible and satisfy  $[\mathbf{\Delta},\mathbf{N}]=[\tilde{\mathbf{\Delta}},\tilde{\mathbf{N}}]=0$, 
$\bd \mathbf{N} = \d(\mathbf{\Delta} \, \mathbf{N})$ and 
$\bd\tilde{\mathbf{N}} = \d( \tilde{\mathbf{N}} \, \tilde{\mathbf{\Delta}})$.
In particular, $\phi'$ is again a solution of (\ref{univ_eq}). 

If $\psi$ and $\tilde{\psi}$ are solutions of (\ref{bdpsi_Delta}) and  (\ref{bdtpsi_Delta}), respectively, and if $\omega,\tilde{\omega}$ satisfy 
\be
 &&   \tilde{\Delta} \, \omega - \omega \, \mathbf{\Delta} 
    = \tilde{\psi} \, \Theta  \, ,  \qquad
      \tilde{\mathbf{\Delta}} \, \tilde{\omega} 
      - \tilde{\omega} \, \Delta 
    = \tilde{\Theta} \, \psi \, ,  \\
 && \bd\omega 
    = (\d\omega) \, \mathbf{\Delta} 
      - (\d\tilde{\Delta}) \, \omega 
      + (\d\tilde{\psi}) \, \Theta
    = \tilde{\Delta} \, \d\omega - \omega \, \d\mathbf{\Delta} 
      - \tilde{\psi} \, \d\Theta \, , \\
 && \bd\tilde{\omega} 
    =(\d\tilde{\omega}) \, \Delta
      - (\d\tilde{\mathbf{\Delta}}) \, \tilde{\omega} 
      + (\d\tilde{\Theta}) \, \psi
    = \tilde{\mathbf{\Delta}} \, \d\tilde{\omega} 
      - \tilde{\omega} \, \d\Delta
      -\tilde{\Theta} \, \d\psi \, , 
\ee
then one verifies by direct calculation that
\be
    \psi' 
 = \psi - \Theta' \, \tilde{\mathbf{N}} \, \tilde{\omega}
 = \psi - \Theta \, \mathbf{\Omega}^{-1} \tilde{\omega} \, , \qquad
   \tilde{\psi}'
 = \tilde{\psi} - \omega \, \mathbf{N} \, \tilde{\Theta}' 
 = \tilde{\psi} - \omega \, \mathbf{\Omega}^{-1} \tilde{\Theta} \, ,
   \label{bDarboux_psi}
\ee
satisfy again (\ref{bdpsi_Delta}), respectively (\ref{bdtpsi_Delta}), 
with $\phi$ replaced by $\phi'$ defined above.  
If $\psi$ is invertible, it is a solution of (\ref{mpd}) and then 
also $\psi'$, if invertible. Correspondingly, $\tilde{\psi}^{-1}$ and 
then also $\tilde{\psi}'^{-1}$ solves  
$\d ( [\bd g - \d(g \tilde{\Delta})] \, g^{-1})=0$.

\subsection{A linearization approach}
\label{subsec:linearization}
Let us consider (\ref{bdphi-psi}) in the form  
\be
    \bd \Phi = (\d \Phi) \, Q \, \Phi + \d \Psi \, ,  \label{bdPhi-Psi}
\ee
where $\d Q = 0$. The reason for the introduction of $Q$ will be given below. 
Setting $\Psi = \Phi \, R$ with a $\d$-constant $R$, this becomes 
\be
    \bd \Phi = (\d \Phi) ( Q \Phi + R ) \, .  \label{bdPhi-G}
\ee
Next we express $\Phi$ as
\be
    \Phi = Y \, X^{-1} \, ,   \label{Phi=YX^-1}
\ee
and impose the constraint
\be
     R X + Q Y = X P    \label{RX+QY=XP}
\ee
with some $P$. Multiplying (\ref{bdPhi-G}) by $X$ 
from the right, leads to
\be
    \bd Y - \Phi \,\bd X = (\d Y) \; P - \Phi \, (\d X) \, P \, ,
\ee
which is a consequence of the two linear equations
\be
    \bd Y = (\d Y) \, P \, , \qquad \bd X = (\d X) \, P \; . \label{XY-lin-eqs}
\ee
The following theorem is now easily verified.\footnote{The proof of the 
theorem does \emph{not} use $\bd^2=0$.}

\begin{theorem}
\label{theorem:lin}
Let $X,Y$ solve the linear equations (\ref{XY-lin-eqs}) and the constraint 
(\ref{RX+QY=XP}) with $\d$-constant $R$ and some $P$, and let 
$X$ be invertible. Then $\Phi = Y \, X^{-1}$ solves 
\be
   \bd \, \d \, \Phi = \d \Phi \, Q \, \d \Phi  \; .   \label{univ_eqQ}
\ee
\end{theorem}
\vskip.1cm

Let $\Phi$ take values in the algebra of $M \times N$ matrices over $\cA$. 
The other objects above are then also matrices with appropriate dimensions. 
If $Q$ has \emph{rank one} over $\cA$, i.e. $Q = V U^T$ with ($\d$- and $\bd$-) 
constant vectors $U,V$ having entries in $\cA$, then  
\be
       \phi = U^T \Phi V
\ee
solves (\ref{univ_eq}) if $\Phi$ solves (\ref{univ_eqQ}). The above theorem 
provides us with a method to construct exact solutions of the nonlinear equation 
(\ref{univ_eqQ}) from solutions of linear equations, and the last argument 
shows how these generate exact solutions of (\ref{univ_eq}). 
Since $M$ and $N$ can be chosen freely, in this way we obtain exact solutions of (\ref{univ_eq}) involving an arbitrarily large number of parameters. This 
partly explains the existence of infinite families of solutions like 
multi-solitons.

More generally, if $Q = V U^T$ with constant $M \times m$ matrix $U$ and 
$N \times m$ matrix $V$, then $\phi = U^T \Phi V$ solves (\ref{univ_eq}) 
in the algebra of $m \times m$ matrices (with entries in $\cA$) if 
$\Phi$ solves (\ref{univ_eqQ}).

A somewhat weaker version of the theorem is obtained 
by extending (\ref{RX+QY=XP}) to  
\be
    H Z = Z P  \qquad \mbox{where} \qquad 
    Z = \left( \begin{array}{c} X \\
                  Y \end{array}\right) \, , \quad
    H = \left(\begin{array}{cc} R & Q \\ S & L
                \end{array}\right) \, ,    \label{HZ=ZP}
\ee
with constant matrices $L$ and $S$. This imposes the additional equation 
$S X + L Y = Y P$ on $X$ and $Y$. Together with (\ref{RX+QY=XP}) it 
implies the algebraic Riccati equation
\be
    S + L \, \Phi - \Phi \, R - \Phi \, Q \, \Phi = 0 \; . 
       \label{algRicc}
\ee
The two equations (\ref{XY-lin-eqs}) 
combine to
\be
      \bd Z = \d Z \, P \; .    \label{Z_lin_sys}
\ee
The equations for $Z$ are form-invariant (with the same $P$) under a transformation 
\be
    Z = \Gamma \, Z' \, , \qquad  
    H = \Gamma \, H' \, \Gamma^{-1} \, ,  \label{Gamma-transf}
\ee
with a ($\d$- and $\bd$-) constant matrix $\Gamma$.  
Such a transformation relates solutions of two versions of (\ref{univ_eqQ}) 
corresponding to two different $Q$'s. One can therefore use the theorem  
with a simple form of $H$ (our choice of $H'$) and then apply a transformation 
to generate a solution associated with a more complicated choice of $H$. 
Choosing
\be
     H' = \left(\begin{array}{cc} R & 0 \\ 0 & L
                \end{array}\right) 
     \quad \mbox{or} \quad 
     H' = \left(\begin{array}{cc} L & I_N \\ 0 & L
                \end{array}\right) \, , \quad
     \mbox{and} \quad
     \Gamma = \left(\begin{array}{cc} I_N & -K \\ 0 & I_M 
                 \end{array}\right)   \label{H-normal-forms}
\ee
($I_N$ is the $N \times N$ unit matrix, $M=N$ in the second case), 
yields the next result. 

\begin{corollary}
\label{cor:Phi-K-formula}
Let $X'$ and $Y'$ solve
\be
    \bd X' = (\d X') \, P \, , \qquad 
    \bd Y' = (\d Y') \, P \, , \qquad  
   L \, Y' = Y' \, P 
\ee
and 
\be
      R \, X' = X' \, P \qquad \mbox{respectively} \qquad 
      L \, X' + Y' = X' \, P \; . 
\ee 
Then $\Phi = Y' \, (X'- K \, Y')^{-1}$ 
with a constant matrix $K$ solves (\ref{univ_eqQ}) with
\be
    Q = R \, K - K \, L  \qquad \mbox{respectively} 
    \qquad  Q = I + [L,K] \; .        \label{Q-K}
\ee
\end{corollary}
\vskip.1cm

\noindent
\textit{Remark.} Solutions of the modified pseudodual equation 
(\ref{mpd}) are obtained as follows. 
Let $G$ be an $m \times N$ matrix solution of 
\be
   \bd (G X) = \d (G X) \, P \qquad \mbox{and} \qquad
   (G X) \, P - \Delta \, (G X) = \mathcal{C} \, X \, ,
\ee
where $\mathcal{C}$ and $\Delta$ satisfy $\d \mathcal{C}=0$ and 
$\bd \Delta = (\d \Delta) \, \Delta$. 
Requiring (\ref{Phi=YX^-1}), (\ref{RX+QY=XP}) and (\ref{XY-lin-eqs}), 
one finds that $G$ solves 
\be
   \bd G + G \, Q \, \d \Phi = \d ( \Delta \, G) \; .
\ee
With $Q$ as specified above, it follows that $g = (G V)^{-1}$, provided 
the inverse exists, solves (\ref{mMiura2}) with $\phi' = U^T \Phi V$. 
As a consequence, $g$ solves (\ref{mpd}). 
\vskip.1cm

\noindent
\textit{Remark.} A relation with the bDT is established as follows. 
If the algebraic Riccati equation (\ref{algRicc}) holds, assuming 
$\d$-constant $L$ and $R$ we have in addition to (\ref{bdPhi-G}) 
also $\bd \Phi = (L - \Phi Q) \, \d \Phi$. 
Now let $Q = V U^T$. Setting $\Theta  = U^T \Phi$ and 
$\tilde{\Theta} = \Phi V$,
we obtain (\ref{bdTheta,bdtTheta}) with $\mathbf{\Delta} = R$ and 
$\tilde{\mathbf{\Delta}} = L$. With the identification 
$\mathbf{\Omega} = \Phi$, we find that (\ref{Theta_conds}) holds 
if $S=0$. 
Furthermore, (\ref{bDT-phi'}) yields $\phi'=0$.

\section{Examples}
\label{section:examples}
\setcounter{equation}{0}
In some examples presented below, the graded algebra will be taken 
of the form $\Omega(\cA) = \cA \otimes_{\mathbb{C}} \bigwedge(\mathbb{C}^n)$ 
where $\bigwedge(\mathbb{C}^n)$ is the exterior algebra of $\mathbb{C}^n$. 
It is then sufficient to define the maps $\d$ and $\bd$ on $\cA$. 
They extend to $\Omega(\cA)$ in an obvious way, treating elements of 
$\bigwedge(\mathbb{C}^n)$ as constants. 
$\xi_1, \ldots, \xi_n$ denotes a basis of $\bigwedge^1(\mathbb{C}^n)$.

\subsection{Self-dual Yang-Mills (sdYM) equation}
Let $\cA$ be the algebra of smooth complex functions of complex variables 
$y,z$ and their complex conjugates $\bar{y}, \bar{z}$. Let
\be
    \d f = \mp f_y \, \xi_1 + f_z \, \xi_2 \, , \qquad
    \bd f = f_{\bar{z}} \, \xi_1 + f_{\bar{y}} \, \xi_2 
\ee
for $f \in \cA$. This determines a BDGA. Then (\ref{univ_eq}), 
for an $m \times m$ matrix $\phi$ with entries in $\cA$, is equivalent to
\be
    \phi_{\bar{y}y} \pm \phi_{\bar{z}z} + [\phi_y , \phi_z] = 0 \, ,
            \label{sdYM-phi}
\ee
which is a potential form of the (Euclidean or split signature) 
sdYM equation (see e.g. \cite{Maso+Wood96}). 
Writing $J$ instead of $g$, the Miura transformation 
(\ref{Miura}) becomes $J_{\bar{y}}J^{-1} = \phi_z$ and 
$J_{\bar{z}}J^{-1} = \mp \phi_y$, and the pseudodual of (\ref{sdYM-phi}) 
takes the form 
\be
   (J_{\bar{y}} J^{-1})_y \pm (J_{\bar{z}} J^{-1})_z = 0 \, , 
\ee
which is another well-known potential form of the sdYM equation. 

Turning to DTs, we choose $\Delta, \tilde{\Delta}$ 
constant. Then (\ref{bdpsi_Delta}) reads
\be
    \psi_{\bar{z}} 
 &=& \mp ( \phi_y \, \psi + \psi_y \, \Delta )
  = J_{\bar{z}} \, J^{-1} \, \psi \mp \psi_y \, \Delta \, , \\
    \psi_{\bar{y}} 
 &=& \phi_z \, \psi + \psi_z \, \Delta 
  = J_{\bar{y}} \, J^{-1} \, \psi + \psi_z \, \Delta \; .
\ee
For an invertible solution $\theta$ of this system, a new solution is given 
by (\ref{Darb_phi'}), respectively 
$J' = (\theta \, \Delta \, \theta^{-1} - C') \, J \, \mathcal{K}^{-1}$ 
via (\ref{phi-g-BT}). (\ref{bdtpsi_Delta}) becomes
\be
 \tilde{\psi}_{\bar{z}} &=& \pm ( \tilde{\psi} \, \phi_y 
               - \tilde{\Delta} \, \tilde{\psi}_y )
  = -\tilde{\psi} \, J_{\bar{z}} \, J^{-1} 
    \mp \tilde{\Delta} \, \tilde{\psi}_y \, ,  \\
 \tilde{\psi}_{\bar{y}} &=& -\tilde{\psi} \, \phi_z 
    + \tilde{\Delta} \, \tilde{\psi}_z
  = - \tilde{\psi} \, J_{\bar{y}} \, J^{-1} 
    + \tilde{\Delta} \, \tilde{\psi}_z \; . 
\ee
(\ref{Omega}) reads 
$\tilde{\Delta} \, \Omega - \Omega \, \Delta = \tilde{\psi} \, \psi$ 
and (\ref{bdOmega}) takes the form
\be
  \Omega_{\bar{z}} = \mp ( \Omega_y \, \Delta + \tilde{\psi}_y \, \psi ) \, , 
                     \qquad
  \Omega_{\bar{y}} = \Omega_z \, \Delta + \tilde{\psi}_z \, \psi \; .
\ee
In this way one recovers corresponding formulae in \cite{NGO00}.
Corollary~\ref{cor:Phi-K-formula} provides a more easily 
applied construction of exact solutions (see also \cite{DMH07disp}).

\subsection{Pseudodual chiral model hierarchy}
Let $\mathfrak{M}$ be a space with coordinates $x_1,x_2,\ldots$. 
On smooth functions on $\mathfrak{M}$ we define
\be
    \d f = \sum_{n\geq 1} f_{x_n} \, \d x_n \, , \qquad
    \bd f = \sum_{n\geq 1} f_{x_{n+1}} \, \d x_n \; . 
\ee
Hence $\d$ is the ordinary exterior derivative. 
Let $\cA$ be the algebra of $m \times m$ matrices of smooth functions and 
$\Omega(\cA) = \cA \otimes_{C^\infty(\mathfrak{M})} \bigwedge(\mathfrak{M})$, 
where $\bigwedge(\mathfrak{M})$ is the algebra of differential forms on $\mathfrak{M}$. 
This determines a BDGA $(\Omega(\cA),\d,\bd)$ and 
(\ref{univ_eq}) reproduces the hierarchy of the $gl(m, \mathbb{C})$ 
`pseudodual chiral model' in $2+1$ dimensions, 
\be
    \phi_{x_{n+1},x_m} - \phi_{x_{m+1},x_n} =
    [\phi_{x_n},  \phi_{x_m}] \, , \qquad \quad m,n=1,2,\ldots  \; .
\ee
The first (non-trivial) equation is a well-known reduction of the 
sdYM equation. $\phi$ can be restricted to any Lie subalgebra of 
$gl(m,\mathbb{C})$, but corresponding conditions then have to be 
imposed on the solution generating methods. 
The method of section~\ref{subsec:linearization} has been 
applied in \cite{DMH07disp}. In the $su(m)$ 
case, a variant of corollary~\ref{cor:Phi-K-formula} has been used 
in particular to construct multiple lump solutions.

\subsection{The potential KP (pKP) equation}
On smooth functions of $x,y,t$ we define
\be
 \d f = [\pa_x,f] \, \xi_1 + \frac{1}{2}[\pa_y+\pa_x^2,f] \, \xi_2 \, , \;
 \bd f = \frac{1}{2} [\pa_y-\pa_x^2,f] \, \xi_1 
         + \frac{1}{3} [\pa_t-\pa_x^3,f] \, \xi_2 \, .
\ee
Besides smooth functions of $x,y,t$ with values in some associative 
algebra, $\cA$ must also contain powers of the partial derivative 
operator $\pa_x$.  
Then (\ref{univ_eq}) becomes the (noncommutative) pKP equation, 
and (\ref{mpd}) with $\Delta = -\pa_x$ the (noncommutative) mKP equation.
Concerning DTs, we set
$\Delta = \Delta' = \tilde{\Delta} = C' = -\pa_x$, and 
$\mathcal{M} = \mathbf{N} = \tilde{\mathbf{N}} = I$.
Then (\ref{Darb_phi'}) and (\ref{Darb_psi'}) take the form 
$\phi' = \phi+\theta_x \theta^{-1}$ and 
$\psi' = \psi_x - \theta_x \theta^{-1} \, \psi$, respectively.
Equation (\ref{bdpsi_Delta}) becomes
\be
    \psi_y = \psi_{xx} + 2 \, \phi_x \psi \, , \qquad
    \psi_t = \psi_{xxx} + 3 \, \phi_x \psi_x 
             + \frac{3}{2} (\phi_y + \phi_{xx}) \, \psi \, , 
\ee
a familiar Lax pair for the pKP equation. The same equations, 
with $\psi$ replaced by $g$, are obtained from (\ref{mMiura}), which 
means that the DT $\psi \mapsto \psi'$ acts on 
mKP solutions. Turning to bDTs, (\ref{bdtpsi_Delta}) reads
\be
    \tilde{\psi}_y = -\tilde{\psi}_{xx} - 2 \, \tilde{\psi} \, \phi_x \, , \qquad
    \tilde{\psi}_t = \tilde{\psi}_{xxx} + 3 \, \tilde{\psi}_x \phi_x 
        - \frac{3}{2} \tilde{\psi} (\phi_y - \phi_{xx}) \; .
\ee
(\ref{Omega}) becomes $\Omega_x = -\tilde{\psi} \, \psi$,  
and (\ref{bdOmega}) yields
\be
 \Omega_y = \tilde{\psi}_x \psi - \tilde{\psi} \, \psi_x \, , \qquad
 \Omega_t = -\tilde{\psi}_{xx} \psi + \tilde{\psi}_x \psi_x
            - \tilde{\psi} \, \psi_{xx} - 3 \, \tilde{\psi} \, \phi_x \psi \; .
\ee
These are well-known formulae, see e.g. \cite{Matv+Sall91,Gils+Nimm07}.

The $\xi_1$-part of (\ref{Z_lin_sys}) is $Z_y-Z_{xx}=2 Z_x (P+\pa_x)$. 
Choosing $P= - I_N \, \pa_x$, this is the heat equation $Z_y=Z_{xx}$. 
The $\xi_2$-part of (\ref{Z_lin_sys}) then becomes the second 
heat hierarchy equation, $Z_t=Z_{xxx}$. Setting 
$R = \tilde{R} - I_N \, \pa_x$ in  (\ref{RX+QY=XP}), turns it into
\be
    X_x = \tilde{R} \, X + Q \, Y  \; . \label{pKP_Xx}
\ee
Now theorem~\ref{theorem:lin} expresses a result for the pKP 
equation \cite{DMH07Burgers,DMH07Wronski} that extends to the whole 
pKP hierarchy, see the next subsection.

\subsection{Kadomtsev-Petviashvili hierarchy}
On smooth functions of variables $x$ and $\bt=(t_1,t_2,\ldots)$ 
we define
\be
  \d f = [\E_\la,f] \, \xi_1 + [\E_\mu,f] \, \xi_2 \, ,  \;
  \bd f = [(\la^{-1}-\pa_x) \E_\la,f] \, \xi_1 
          + [(\mu^{-1}-\pa_x) \E_\mu,f] \, \xi_2 \;
\ee
where $\E_\la$ is the Miwa shift operator with an indeterminate $\lambda$, i.e. 
$\E_\la f = f_{[\la]} \E_\la$ where $f_{\pm [\la]}(x,\bt) = f(x,\bt \pm [\la])$ 
with $[\la] = (\la, \la^2/2, \la^3/3,\ldots)$. Furthermore, $\pa_x$ is the 
partial derivative operator with respect to $x$. Let $\cA$ contain the 
algebra of $m \times m$ matrices of smooth functions. The above expressions 
for $\d f, \bd f$ require that $\cA$ also contains the Miwa shift operators 
and powers of $\pa_x$. 
(\ref{univ_eq}) is equivalent to the following functional representation 
\cite{Bogd+Kono98,DMH06func} of the (matrix) potential KP hierarchy, 
\be
     (\phi_{-[\la]}-\phi_{-[\mu]})_x 
 &=& (\mu^{-1}-\phi+\phi_{-[\mu]})_{-[\la]}(\la^{-1}-\phi+\phi_{-[\la]}) 
     \nonumber \\
 && - (\la^{-1}-\phi+\phi_{-[\la]})_{-[\mu]}(\mu^{-1}-\phi+\phi_{-[\mu]}) \; .
\ee
In particular, $\phi_{t_1} = \phi_x$. 
The linear system (\ref{XY-lin-eqs}), which is (\ref{Z_lin_sys}), takes the form
\be
   (Z - Z_{-[\la]}) (P + \pa_x - \la^{-1}) + Z_x = 0  \; .
\ee
Choosing $P = -I_N \, \pa_x$ and applying a Miwa shift, this reduces to 
\be
    \la^{-1} (Z - Z_{-[\la]}) = Z_x \, , 
\ee
which is the linear heat hierarchy $Z_{t_n} = \pa_x^n(Z)$, $n=2,3,\ldots$. 
Choosing moreover $R = \tilde{R} - I_N \, \pa_x$, (\ref{RX+QY=XP}) takes 
the form (\ref{pKP_Xx}). 
Now theorem~\ref{theorem:lin} reproduces theorem~4.1 in \cite{DMH07Burgers}. 
See also \cite{DMH07Ricc,DMH07Wronski,DMH07def} for exact solutions 
obtained in this way.

\subsection{2-dimensional Toda lattice (2dTL) equation}
\label{subsec:Toda}
On smooth functions of $x,y$ and an additional discrete variable, we set
\be
    \d f = [\Lambda \, , \, f] \; \xi_1
             + [\pa_y \, , \, f] \; \xi_2 \, , \qquad
   \bd f = [\pa_x \, , \, f] \; \xi_1
             - [\Lambda^{-1} \, , \, f] \; \xi_2 \, ,
\ee
where $\Lambda$ is the shift operator in the discrete variable.
$\cA$ must also contain powers of $\Lambda$.
Now (\ref{univ_eq}) leads to the noncommutative 2dTL equation
\be
   \tilde{\phi}_{xy}
 = ( \tilde{\phi}^+ - \tilde{\phi}) ( I + \tilde{\phi}_y )
   - ( I + \tilde{\phi}_y ) ( \tilde{\phi} - \tilde{\phi}^- )
  \qquad \mbox{where} \quad \tilde{\phi} := \phi \, \Lambda
      \label{nc2dToda}
\ee
and $\tilde{\phi}^+ = \Lambda(\tilde{\phi})$, $\tilde{\phi}^- = \Lambda^{-1}(\tilde{\phi})$.
For a commutative algebra $\cA$, this takes the form
\be
    (\log(1+v))_{xy} = v^+ - 2 \, v + v^-
\ee
in terms of $v=\tilde{\phi}_y$. Choosing $\Delta = \nu \, \Lambda^{-1}$ with 
a constant $\nu$, (\ref{mpd}) takes the form
\be
    ( g_x \, g^{-1} )_y
 = g^+ \, g^{-1} \, ( I - \nu \, g_y \, g^{-1} )
   - ( I - \nu \, g_y \, g^{-1} ) \, g \, (g^-)^{-1} \; .
     \label{ncmToda}
\ee
If $g=e^q$ with a scalar function $q$, this is the
\emph{modified Toda equation} \cite{Hiro04}
\be
    q_{x y} = (1-\nu \, q_y) ( e^{q^+ -q} - e^{q-q^-} ) \; .
\ee

Inspection of (\ref{bdpsi_Delta}) suggests $\Delta = \nu \, \Lambda^{-1}$, 
which turns it into 
$\psi_x = (\tilde{\phi}^+ - \tilde{\phi}) \, \psi + \nu \, (\psi^+ - \psi)$ 
and $\nu \,\psi_y = -  \tilde{\phi}_y \, \psi^- +  (\psi - \psi^-)$.
Setting $\mathcal{M} = \mathcal{M}' = \Delta'= \Lambda^{-1}$ 
and $C'=0$ in section~\ref{subsec:Darboux}, 
(\ref{Darb_phi'}) and (\ref{Darb_psi'}) yield the familiar
DT $\tilde{\phi}' = \tilde{\phi} + \theta \, (\theta^-)^{-1}$,
$\psi' = \nu \, \psi - \theta \, (\theta^-)^{-1} \, \psi^-$.
For the iterated transformation (with 
$\Delta_k = \mathcal{M}_k = \Lambda^{-1}$), we obtain in 
quasideterminant notation (see also \cite{EGR97,Li+Nimm07}) 
\be
     \tilde{\phi}_{[N+1]}
 &=& \tilde{\phi}
     + \sum_{k=1}^N \theta_{[k]} \, (\theta_{[k]}^-)^{-1} \quad
     \mbox{where} \quad
     \theta_{[k+1]} = \left| \begin{array}{ccc}
   \theta_1 & \cdots & \boxed{\theta_{k+1}} \\[.2cm]
   \theta_1^- & \cdots & \theta_{k+1}^-  \\
   \vdots & \ddots & \vdots  \\
   \theta_1^{(k-)} & \cdots & \theta_{k+1}^{(k-)}
   \end{array} \right| \, , \nonumber \\
     \psi_{[N+1]}
 &=& \left| \begin{array}{cccc}
   \theta_1 & \cdots & \theta_N & \boxed{\nu^N \psi} \\[.2cm]
   \theta_1^- & \cdots & \theta_N^- & \nu^{N-1}\psi^- \\
   \vdots & \ddots & \vdots & \vdots \\
   \theta_1^{(N-)} & \cdots & \theta_N^{(N-)} & \psi^{(N-)}
   \end{array} \right| \; .
\ee
If $\psi_{[N+1]}$ is invertible, then it solves the noncommutative 
(or non-Abelian) modified 2dTL equation (\ref{ncmToda}), and for 
$\nu=0$ it solves the ordinary noncommutative 2dTL equation.
In a similar way, one recovers the bDT in \cite{Li+Nimm07}.
In the approach of section~\ref{subsec:linearization}, setting
$X = \Lambda \, \tilde{X}$, $R = \tilde{R} \, \Lambda^{-1}$,
$P = \Lambda^{-1}$, we have
\be
   \tilde{\Phi} = \Phi \, \Lambda = Y \, \tilde{X}^{-1} \, , \quad
   \tilde{Z}_x = \tilde{Z}^+ - \tilde{Z} \, , \quad
   \tilde{Z}_y = \tilde{Z}^- - \tilde{Z} \, , \quad
   \tilde{Z} = \left(\begin{array}{c} \tilde{X} \\ Y \end{array}\right)
   \, ,
\ee
and $\tilde{R} \, \tilde{X} + Q \, Y = \tilde{X}^+$.
Extending the last relation to $H \, \tilde{Z} = \tilde{Z}^+$,
we obtain
\be
 \tilde{Z}_n = e^{x \, (H-I_{N+M}) + y \, (H^{-1}-I_{N+M}) } \,H^n \tilde{Z}_0 ,
\ee
assuming $H$ invertible. Using (\ref{Gamma-transf}) with
(\ref{H-normal-forms}), and imposing $\mathrm{rank}(Q)=m$ on $Q$
then given by (\ref{Q-K}), one can now compute explicit solutions
of (\ref{nc2dToda}) in the algebra of $m \times m$ matrices.

\subsection{Lotka-Volterra (LV) lattice equation}
Let 
\be
    \d f = [\Lambda^2,f] \, \xi_1 + [\Lambda,f] \, \xi_2 \, , \quad
    \bd f = \dot{f} \, \xi_1 + [\Lambda^{-1},f] \, \xi_2 \, , 
\ee
with the shift operator $\Lambda$ and $\dot{f}=f_t$. Introducing 
$a = \varphi^+ - \varphi -I$ where $\varphi = \phi \Lambda^2$, 
(\ref{univ_eq}) becomes the (noncommutative) LV lattice equation
\be
    \dot{a} = a^+ a - a \, a^- \; .
\ee
In terms of $b = g^{+} g^{-1}$ and $c = \dot{g} \, g^{-1}$, the modified  
Miura transformation (\ref{mMiura}) with $\Delta = \la \Lambda^{-2}$ reads
\be
   a = - b^{-} - \la \, (b \, b^{-} - b^{-} ) \, , \qquad
   c = 2 - \la + (\la-1)(b + b^{-}) - \la \, b b^{-} \; .
\ee
Since $\dot{b} = c^{+} b - b \, c$ by definition of $b$ and $c$, 
we obtain 
\be
   \dot{b} = (\la-1)(b^{+} b - b \, b^{-}) - \la \, (b^{+} b^2 - b^2 b^{-}) \, ,
\ee
a noncommutative version of the \emph{modified LV lattice} \cite{Suri99}. 
For a DT, appropriate choices are 
$\Delta = \la \Lambda^{-2}$ and $\mathcal{M} = -\Delta^{-1}$. 
In the approach of section~\ref{subsec:linearization}, we set
\be
    X = \Lambda^2 \, \tilde{X} \, , \qquad 
    R = \tilde{R} \, \Lambda^{-2} \, , \qquad
    P = -\tilde{P}^{-1} \, \Lambda^{-2} \, ,
\ee
with constant $\tilde{P}$. 
Then (\ref{RX+QY=XP}) takes the form
\be
  \tilde{R} \, \tilde{X} + Q \, Y = -\tilde{X}^{++} \, \tilde{P}^{-1} \, , 
       \label{LV_algcond}
\ee
and (\ref{XY-lin-eqs}) leads to 
\be
    \dot{\tilde{X}}= -(\tilde{X}^{++}-\tilde{X})\tilde{P}^{-1}, \qquad
    (\tilde{X}^+ - \tilde{X}) = (\tilde{X}^+ - \tilde{X})^-\tilde{P} \, , 
\ee
and the same equations for $Y$. These equations are solved by
\be
  \tilde{X}_n = A + B \tilde{P}^n e^{t(\tilde{P}^{-1}-\tilde{P})} 
                \, , \qquad
  Y_n = C + D \tilde{P}^n e^{t(\tilde{P}^{-1}-\tilde{P})} \, ,
\ee
with constant matrices $A,B,C,D$. Inserting this 
in (\ref{LV_algcond}), we find the constraints
\be
    Q C + \tilde{R} A = - A \tilde{P}^{-1} \, , \qquad 
    Q D + \tilde{R} B = - B \tilde{P} \; .
\ee
If $Q=V U^t$, then via $\varphi_n = U^t Y_n \tilde{X}_n^{-1} V$ and 
$a_n = \varphi_{n+1} - \varphi_n - 1$ we obtain solutions of 
the scalar LV lattice equation. 
An extension of (\ref{LV_algcond}) in the sense of (\ref{HZ=ZP}) 
turns out to be too restrictive.

\section{Final remarks}
\label{section:conclusions}
Any BDGA formulation of an integrable system 
provides us with a Lax pair which is linear in the spectral parameter, 
a situation well known from the (anti-) self-dual Yang-Mills system. 
The restriction to a special form of Lax pairs is what enabled us 
to work out calculations, that appeared in the literature for specific 
integrable systems, in a universal way. Although many integrable models 
indeed fit into this framework, it is not clear to what extent this covers 
the existing variety of integrable systems. If some integrable system 
possesses a Lax pair that is non-linear in the spectral 
parameter, this does \emph{not} exclude the existence of a Lax pair 
\emph{linear} in such a parameter (see \cite{Bord+Yano95}, 
for example).

\end{document}